\documentstyle[12pt]{article}
\begin{document}

\centerline{\large LINEAR RESPONSE THEORY IN STOCHASTIC RESONANCE}

\vskip 0.2truecm
\centerline {M.I. Dykman\footnote{Permanent address: Department of Physics,
Stanford University, Stanford, California 94305, USA}}
\centerline {School of Physics and Materials, Lancaster University,
Lancaster LA1 4YB, UK}

\centerline {H. Haken}
\centerline {Institut f\"{u}r Theoretische Physik und Synergetik,
Universit\"{a}t Stuttgart,}
\centerline {D-70550 Stuttgart, Germany}

\centerline {Gang Hu}
\centerline {Department of Physics, Beijing Normal University, Beijing 100875,
China}

\centerline {D.G. Luchinsky\footnote{Permanent address: Institute for
Metrology, 117965 Moscow, Russia}, R. Mannella\footnote{Permanent Address:
Dipartimento di Fisica, Universit\`{a} di Pisa, 56100 Pisa, Italy},
P.V.E. McClintock}
\centerline {School of Physics and Materials, Lancaster University,
Lancaster LA1 4YB, UK}

\centerline {C.Z. Ning}
\centerline {Institut f\"{u}r Theoretische Physik und Synergetik,
Universit\"{a}t Stuttgart,}
\centerline {W-7000 Stuttgart 80, Germany}

\centerline {N.D. Stein}
\centerline {School of Physics and Materials, Lancaster University,
Lancaster LA1 4YB, UK}

\centerline {N.G. Stocks}
\centerline {Department of Engineering, University of Warwick, Coventry
CV4 7AL, UK.}

\vskip 0.2truecm
\noindent
\underbar{Abstract}

The susceptibility of an overdamped Markov system fluctuating in a bistable
potential of general form is obtained by analytic solution of the Fokker-Planck
equation (FPE) for low noise intensities.  The results are discussed
in the context of the LRT theory of stochastic resonance.  They go over into
recent results (Gang Hu et al {\em Phys. Lett. A} {\bf 172}, 21, 1992)
obtained from the FPE for the case of a symmetrical potential, and they
coincide with the LRT results (Dykman et al, {\em Phys. Rev. Lett.} {\bf 65},
2606, 1990;  {\em JETP Lett} {\bf 52}, 144, 1990; {\em Phys. Rev. Lett.}
{\bf 68}, 2985, 1992) obtained for the general case of bistable systems.


\newpage

The phenomenon of stochastic resonance (SR), where a periodic signal in a
system and the signal-to-noise ratio are enhanced by external noise,
has attracted much attention recently and by now has been observed in
several physical systems (see [1] and references therein).  The occurrence
of SR is related to the nonlinearity of a system.  At the same time, for
small enough amplitude $A$ of the periodic force causing a signal, the
amplitude of the signal is proportional to $A$, and therefore many aspects
of SR can conveniently and effectively be analysed [2-6] in terms of standard
linear response theory (LRT) [7]. The LRT analysis amounts to finding the
susceptibility  of the system $\chi (\Omega)$ and its dependence on the noise
intensity $D$ and on the frequency $\Omega$ of the force. The susceptibility
is defined in a standard way [7] as the coefficient of proportionality
between the periodic term in the ensemble-averaged value of the coordinate
$\delta \langle q(t)\rangle$ (the signal) and the force $A\exp (-i\Omega t)$,

\begin{equation}
\delta \langle q(t)\rangle = \chi (\Omega) A\exp (-i\Omega t)
\end{equation}

\noindent
(in the case of a nonstationary system the definition (1) has to be
modified). The amplitude
of the signal is equal to $|\chi (\Omega)| A$, and SR arises when $|\chi
(\Omega)|$ increases
with $D$ in some interval of $D$ (and then decreases again for higher $D$,
so that the dependence
on noise intensity is of the form of a resonant curve, thus justifying the term
\lq\lq stochastic resonance").

The LRT formulation is not confined to particular systems or particular
types of noise
and, indeed, was used to predict and observe SR [4, 6] in systems
drastically different from
that performing Brownian motion in a symmetric double-well potential which is
usually considered as the prototype of SR-displaying systems [1]. One of
the advantageous
features of this formulation is that, in the case of systems at thermal
equilibrium or quasi-equilibrium
(in which case it is temperature $T$ that corresponds to the noise
intensity $D$), the
susceptibility $\chi (\Omega)$ is related, via the fluctuation-dissipation
relations, to the
spectral density of fluctuations (SDF) of the coordinate $q(t)$ [7].
Therefore, if the SDF
is known from the experimental data, it is straightforward to predict
whether the system
will or will not display SR when driven by a force of a given frequency
even without
establishing the underlying physical mechanism; we note that the
corresponding analysis does not assume that the system is Markovian or
whatever - the {\em only}
thing assumed is that it is in thermal equilibrium.

The theoretical evaluation of the susceptibility for some simple model
noise-driven systems
that display SR, and for systems performing Brownian motion in a double-well
potential in particular, has been done in Refs. [2 - 6]. The analysis of
bistable systems
could be done for small noise intensities where there are two strongly
different time
scales that characterize the motion of a system: the relaxation time $t_r$
over which the
system approaches a stable state (the longest of these times), and the
reciprocal probabilities
$W_{nm}^{-1}$ of fluctuational transitions $n\rightarrow m$ ($n,m = 1,2$)
between the states,

\begin{equation}
Wt_r \ll 1, \quad W = W_{12} + W_{21}
\end{equation}

\noindent
Two approaches have been used. The first of these [2, 3] is based on the
idea that, for low
noise intensities, a system spends most of its time localized close to the
stable states, and only occasionally do
there occur fluctuational transitions between the states; therefore the
susceptibility
is a sum of the partial susceptibilities $\chi_n(\Omega)$ from small-amplitude
vibrations about the stable positions $q_n$ ($n = 1,2$), weighted with the
populations
$w_n$ of the stable states, and of the susceptibility related to the
fluctuational
transitions, $\chi_{tr}(\Omega)$, which describes the effect of the
modulation of
the transition probabilities and hence of the populations of the stable
states by a
periodic force (cf. [8]):

\begin{equation}
\chi (\Omega) = \sum_{n = 1,2}w_n\chi_n(\Omega) + \chi_{tr}(\Omega),\quad
w_n=W_{mn}/W
\end{equation}

\noindent
The evaluation of $\chi_n$ reduces to the solution of a linearized problem,
and the
explicit expression for $\chi_{tr}$ has been obtained for systems
driven by Gaussian noise in the range $\Omega t_r \ll 1$ ($\chi_{tr}
\propto W/\Omega$
for $W \ll \Omega$, cf. Eq.(18) below, and it becomes very small for
$\Omega t_r \sim 1$ in the range (2)).

A different approach has been suggested in Ref.5 for the analysis of overdamped
Brownian motion in a symmetrical bistable potential $U(q)$,

\begin{equation}
\dot{q} + U'(q) = f(t), \quad \langle f(t)f(t')\rangle = 2D\delta (t - t')
\end{equation}

\noindent
The basic idea of this approach is to reduce the Fokker-Planck equation

\begin{equation}
\frac{\partial P(q,t)}{\partial t} = \frac{\partial (U'(q)P(q,t))}{\partial q}
+
D\frac{\partial^2 P(q,t)}{\partial q^2}
\end{equation}

\noindent
that corresponds to the Langevin equation (4) to an eigenvalue problem and to
express the susceptibility in terms of the matrix elements of the coordinate
$q$ on the corresponding eigenfunctions. It was then possible to evaluate the
susceptibility explicitly in the range of noise intensities $D$ where (2)
holds.  A systematic perturbation theory in the amplitude of a sinusoidal
force has been developed for systems described by the Fokker-Planck
equation and applied to SR in an effectively two-state approximation in
Ref [9].

The purpose of the present paper is to demonstrate that the two approaches
[2, 3] and [5] give identical results and to generalize the approach [5] to
the case of an arbitrary double-well potential $U(q)$.

The reduction of Eq.(5) to a Hermitian eigenvalue problem is standard
[10, 11]: one seeks the solution in the form

\begin{equation}
P(q,t) = Z^{-1/2}\exp (-U(q)/2D)\sum_{j = 0}^{\infty}|j\rangle\exp
(-\lambda(j)t),
\quad |j\rangle \equiv \psi(j;\,q),
\end{equation}
$$ Z = \int_{-\infty}^{\infty} \exp (-U(q)/D)dq$$

\noindent
and then arrives at a Schrodinger-type equation for the eigenfunctions
$\psi(j;\,q)$:

\begin{equation}
-D^2\psi''(j;\,q) + V(q)\psi(j;\,q) = D\lambda(j)\psi(j;\,q), \quad V(q) =
\frac{1}{4}\left( U'\left( q\right)\right)^2 - \frac{1}{2}DU''(q)
\end{equation}

\noindent
The functions $\psi(j;\,q)$ are orthonormal, $\langle j|j'\rangle =
\delta_{jj'}$ (the form of the
expansion (6) is a bit different from that used in [5]). Following the
arguments given in Ref. 5
one can arrive at the following expression for the susceptibility of the
system (4) with
respect to a force $A\exp (-i\Omega t)$:

\begin{equation}
\chi (\Omega) = -2\sum_j(\lambda(j) - i\Omega)^{-1}\langle 0|q|j\rangle
\langle j|\frac{\partial}{\partial q}|0\rangle \quad
\left( \langle j|\hat{L}|j'\rangle \equiv \int_{-\infty}^{\infty}dq\,\psi
(j;\,q)\hat{L}\psi (j';\,q)\right)
\end{equation}

\noindent
The eigenvalues $\lambda(j)$ and the eigenfunctions $\psi(j;\,q)$  for low
noise intensities
$D$ were analysed in Ref. [11]. The qualitative results can be easily
understood by
noticing that, for a bistable initial potential $U(q)$ and small $D$,
the potential $V(q)$ that appears in Eq. (7)
has 3 minima of nearly the same depth
placed close to the stable positions $q_{1,2}$ and an unstable stationary
point $q_s$ (local maximum of $U(q);\;U'(q_{1,2}) = U'(q_s) = 0, \;
U''(q_{1,2}) > 0, \;
U''(q_s) < 0$). Close to the minima $V(q)$ is parabolic:

\begin{equation}
V(q) \approx -\frac{1}{2}DU''_n + \frac{1}{4}(U''_n)^2(q - q_n)^2, \quad |q
- q_n| \ll |q_n - q_s|,
\quad U''_n \equiv U''(q_n)
\end{equation}

$$V(q) \approx \frac{1}{2}D|U''_s| + \frac{1}{4}(U''_s)^2(q - q_s)^2, \quad
|q - q_s|
\ll |q_{1,2} - q_s|, \quad U''_s \equiv U''(q_s)$$

\noindent
The lowest \lq\lq energy levels" $\overline{\lambda_n}(0)$ of the motion within
the wells $n = 1,2$ of the potential $V$ (the overbar is used to indicate
intrawell states , and the subscript enumerates
the wells at the corresponding stable positions $q_n$)
are degenerate, with $\overline{\lambda}_{1,2}(0) = 0$ to the lowest order
in $D$, giving rise to
tunnelling splitting of these levels [11]. The low-lying
excited intrawell states $\overline{|j\rangle _n} \equiv \
\overline {\psi_n}(j;\,q)$ are those of quantum harmonic oscillators, and

\begin{equation}
\overline{\lambda_n}(k) \approx kU''_n
\end{equation}

\noindent
(and of course, high excited states cannot be prescribed to a particular well).

The exact wave function of the ground state (that allows for the tunnelling
splitting) is known
to be

\begin{equation}
\psi (0;\,q) = Z^{-1/2}\exp (-U(q)/2D), \quad \lambda (0) = 0
\end{equation}

\noindent
It has sharp maxima at $q_{1,2}$ (of width $\delta q \sim
(D/U''(q_n))^{1/2}$) and
a minimum at $q_s$. Near the maxima, the function (11) is close to the
intrawell
ground-state oscillatory functions
$\overline{\psi_n}(0;\,q)$ weighted with the factors
$Z^{-1/2}(2\pi D/U''_n)^{1/4}\,\times\exp (-U(q_n)/2D)$.
This makes it possible to evaluate the contribution to the susceptibility
(8) from the
excited intrawell states, $\chi_{intra}(\Omega)$. Since, for a harmonic
oscillator,
the only nonzero matrix elements
of the operator $\partial/\partial q$ are those between adjacent states,
only the first excited intrawell states, $\psi (j;\,q) \approx
\overline{\psi_n}(1;\,q)$ ($n = 1,2$)
will contribute to $\chi_{intra}(\Omega)$.
 Allowing for the explicit form of the matrix elements, known from
quantum mechanics [12], and for Eqs.(9), (10) we get,
with an accuracy to small corrections $\sim D$:

\begin{equation}
\chi_{intra}(\Omega) \approx \sum_{n = 1,2}w_n(U''_n - i\Omega)^{-1}
\end{equation}

\noindent
Here, $w_n$ is the population of the $n$th stable state,

\begin{equation}
w_n = \frac{\exp (-\Delta U_{3 - n}/D)(U''_{3 - n})^{1/2}}{\sum_{m =
1,2}\exp (-\Delta U_{m}/D)
(U''_m)^{1/2}}, \quad \Delta U_m \equiv U(q_s) - U(q_m)
\end{equation}

\noindent
and we have evaluated the integral (6) for $Z$ by the steepest descent method.
The overlap integral of the function $\psi (0;\,q)$ with (non-oscillatory)
wave functions of
highly excited states, as well as with the wave functions that correspond to
the
low-lying states in the well of the potential $V(q)$ centered at $q_s$, are
exponentially
small, and therefore the contribution from the corresponding matrix elements to
the susceptibility (8) can be ignored.

The only other important contribution to $\chi (\Omega)$ besides
(12) is that from the matrix elements calculated between the function (11)
and the function
$\psi (1;\,q)$ corresponding to the excited state that results from the
tunnel splitting
of the degenerate intrawell state, whence

\begin{equation}
\chi (\Omega) \approx \chi_{intra}(\Omega) + \chi_{tun}(\Omega), \quad
\chi_{tun}(\Omega) =
\frac{\lambda(1)}{D}\,
\frac{|\langle 0|q|1\rangle |^2}{\lambda (1) - i\Omega}
\end{equation}

\noindent
Here, we have implemented the interrelation [11], well-known from quantum
mechanics, between the matrix elements
of the operators $\partial/\partial q$ and $q$ in terms of the energy
difference between the states. The problem of evaluating $\chi_{tun}$ comes
to the evaluation of the matrix elements
of the coordinate. The latter
can be done by making use of the explicit form of the wave functions
$|0\rangle$ (11) and
$|1\rangle$. The function $|1\rangle$ as well as the value of $\lambda (1)$
were found in [11].
An alternative way to find them is based on applying to
the present problem  the approach suggested in [12]
to calculate tunnelling level splitting in a symmetrical
double-well potential. It is easy to see that, in the spirit of [12],
the wave function $\psi (1;\,q)$ can be sought in the form

\begin{equation}
\psi (1;\,q) = A_0\psi (0;\,q) + A_1\tilde{\psi}(1;\,q)
\end{equation}

$$\tilde{\psi}(1;\,q) \approx \left\{ \begin{array}{cl}
-\psi (0;\,q), & q_s - q \gg (D/|U''_s|)^{1/2}\\
\psi (0;\,q)(2|U''_s|/\pi D)^{1/2}\int_{q_s}^q\,dq'\,
\exp\left[\frac{U(q') - U(q_s)}{D}\right],
& |q - q_s| \stackrel{<}{\sim}(D/|U''_s|)^{1/2}\\
\psi (0;\,q), & q - q_s \gg (D/|U''_s|)^{1/2}
\end{array}
\right. $$

\noindent
The coefficients $A_{0,1}$ can be obtained from the conditions of
normalization of
$\psi (1;\,q)$ and of orthogonality of
$\psi (1;\,q)$ and $\psi (0;\,q)$, whereas the eigenvalue $\lambda_1$ is
given by
the expression

\begin{equation}
\lambda (1) = - D\psi (0;\,q_s)\left( \frac{d\psi (1;\,q)}{dq}\right)_{q_s}
\left[ \int_{-\infty}^{q_s} dq\,\psi (0;\,q)\psi (1;\,q)\right]^{-1}
\end{equation}

\noindent
that immediately follows from (7), (11).

It is seen from Eqs. (8), (14) - (16) that the susceptibility
$\chi _{tun}(\Omega)$ due to the transitions to the tunnel-split state
$0 \rightarrow 1$ is of the following form

\begin{equation}
\chi _{tun}(\Omega) = w_1w_2(q_2 - q_1)^2\frac{\lambda (1)}{D}\,(\lambda
(1) - i\Omega)^{-1}
\end{equation}

$$\lambda (1) = \pi^{-1}\sum_{n = 1,2}\left[\sqrt{U''_n|U''_s|}\exp
(-\Delta U_n/D)\right]$$

\noindent
In the particular case of a symmetrical potential $U(q)$ Eqs.(1), (12), (14),
(16) go over into
the result obtained in [5].

It is interesting to compare the result of the present calculations with
the expression
(3) applied to the model (4). The explicit form of the susceptibility for
this model
was discussed in [2, 3]. The intrawell susceptibilities
$\chi_n(\Omega) = (U''_n - i\Omega)^{-1}$ follow from Eq.(4)
linearized about the stable positions $q_{1,2}$. The expression
for the susceptibility due to interwell transitions is of the form:

\begin{equation}
\chi_{tr}(\Omega) = w_1w_2(q_2 - q_1)^2\frac{W}{D}\,(W - i\Omega)^{-1}
\end{equation}

\noindent
where $w_n$ is the population of the $n$th state, and $W$ is the sum of the
probabilities of the transitions between the states (cf. (2)). Allowing for
the explicit
expressions for the transition probabilities, one finds that the intrawell
contribution
$\chi_{intra}(\Omega)$ obtained above is precisely equal to that given by
(3), and that
$\lambda (1) = W$ (cf. [10, 11]) and therefore
$\chi_{tun}(\Omega) = \chi_{tr}(\Omega)$, and the susceptibilities as a
whole coincide with each other.
So they do also in the particular
case of a symmetric double-well potential discussed in [5]. The explicit
expression
for the spectral density of fluctuations considered in [5] coincides with
that discussed
in [2,3] for the model (3) as well. We notice that, in the general case of
an asymmetric potential,
the intensity of the transition- (or, equivalently) tunnelling-induced peak
in the
susceptibility (17), (18) is proportional to $w_1w_2 \propto \exp (-|U(q_1)
- U(q_2)|/D)$, and it
drops down
extremely sharply with the difference in the depths of the wells.

It follows from the above discussion that two, seemingly different, approaches
to the analysis of the susceptibility of a double-well system, one
based on a simple physical picture of the motion and the other based on the
solution
of the Fokker-Planck equation when the latter applies, give identical
results.  This
can be regarded as an extra
indication of the usefulness of linear response theory in the context of
stochastic resonance.

\newpage
\centerline {REFERENCES}

\begin{itemize}

\item[1.] Special issue of the {\it J. Stat. Phys.} {\bf 70} (1993) no.1/2.

\item[2.] M. I. Dykman, R. Mannella, P. V. E. McClintock, and N. G. Stocks,
{\it Phys. Rev. Lett.} {\bf 65} (1990) 2606; M. I. Dykman, P. V. E. McClintock,
R. Mannella, and N. G. Stocks, {\it Soviet Phys. JETP Lett.} {\bf 52} (1990)
144.

\item[3.] M. I. Dykman, R. Mannella, P. V. E. McClintock, and
N. G. Stocks, {\it Phys. Rev. Lett.} {\bf 68} (1992) 2985.

\item[4.] N.G. Stocks, N.D. Stein, S.M. Soskin, and P.V.E. McClintock,
{\it J. Phys. A} {\bf 25} (1992) L1119.

\item[5.] G. Hu, H. Haken, and C.Z. Ning, {\it Phys. Lett. A} {\bf 172}
(1992) 21.

\item[6.] M.I. Dykman, D.G. Luchinsky, R. Mannella, P.V.E. McClintock, N.D.
Stein,
and N.G. Stocks, {\it J. Stat. Phys.} {\bf 70} (1993) 463; {\it ibid.},
{\bf 70} (1993) 479.

\item[7.] L.D. Landau and E.M.Lifshitz, {\it Statistical Physics}, 3rd
edition (Pergamon,
Oxford, 1980).

\item[8.] M.I.Dykman and M.A. Krivoglaz, {\it Sov. Phys. JETP} {\bf 50}
(1979) 30;
M.I. Dykman and M.A. Krivoglaz, in {\it Soviet Physics Reviews}, ed. by
I.M. Khalatnikov
(Harwood, New York 1984), Vol. 5, p.265.

\item[9.] Hu Gang, G. Nicolis and C. Nicolis, {\it Phys. Rev. A}
{\bf 42}, 2030 (1990).

\item[10.] N.G. van Kampen, {\it Stochastic Processes in Physics and
Chemistry} (Elsevier,
Amsterdam 1990); H. Risken, {\it The Fokker-Planck Equation},
2nd edition (Springer-Verlag, Berlin 1989).

\item[11.] B. Caroli, C. Caroli, and B. Roulet, {\it J. Stat. Phys.} {\bf
21} (1979) 415.

\item[12.] L.D. Landau and E.M. Lifshitz, {\it Quantum Mechanics}, 3rd edition
(Pergamon, Oxford 1977).

\end{itemize}
\end{document}